\documentclass[%
prl,twocolumn,
superscriptaddress,
showpacs,
 amsmath,amssymb,
 aps,
floatfix,
]{revtex4-1}
 
\usepackage{graphicx}
\usepackage{dcolumn} 
\usepackage{bm}      
\usepackage{subfigure}
\usepackage{amsfonts}
\usepackage[usenames]{color}
\usepackage{rotating}
\usepackage{fontenc}
\usepackage{cancel}
\usepackage{amsthm}
\usepackage{hyperref}
\usepackage[normalem]{ulem}
\usepackage{bibunits}

\defaultbibliography{<bib-file>}
\defaultbibliographystyle{<preferred bib style>}

\definecolor{darkred}{RGB}{139,0,0}
\definecolor{chartreuse}{RGB}{127,255,0}
\definecolor{goldenrod}{RGB}{218,165,32}
\definecolor{gray}{RGB}{127,127,127}

\definecolor{Magenta}{RGB}{255, 0,255}
\definecolor{Orange}{RGB}{255,165, 0}
\definecolor{Gray}{RGB}{127,127,127}

\newcommand{\be}{\begin{equation}}
\newcommand{\ee}{\end{equation}}
\newcommand{\bea}{\begin{eqnarray}}
\newcommand{\eea}{\end{eqnarray}}
\newcommand{\bw}{\begin{widetext}}
\newcommand{\ew}{\end{widetext}}
\newcommand{\mm}{\mathrm}

\newcommand{\bi}{\begin{itemize}}
\newcommand{\ei}{\end{itemize}}

\begin{document}
\begin{bibunit}

\title{Bridges in Complex Networks}

\author{Ang-Kun Wu}
\affiliation{Channing Division of Network Medicine, Brigham and Women's Hospital, Harvard Medical School, Boston, Massachusetts, 02115, USA}
\affiliation{Department of Physics, Chu Kochen Honors College, Zhejiang University, Hangzhou, Zhejiang, 310027, China}

\author{Liang Tian}
\affiliation{Channing Division of Network Medicine, Brigham and Women's Hospital, Harvard Medical School, Boston, Massachusetts, 02115, USA}
\affiliation{College of Science, Nanjing University of Aeronautics and Astronautics, Nanjing 210016, China}

\author{Yang-Yu Liu}
\email[Corresponding author: yyl@channing.harvard.edu]{}
\affiliation{Channing Division of Network Medicine, Brigham and Women's Hospital, Harvard Medical School, Boston, Massachusetts, 02115, USA}
\affiliation{Center for Cancer Systems Biology, Dana-Farber Cancer Institute, Boston, Massachusetts, 02115, USA}

\date{\today}

\begin{abstract}
A bridge in a graph is an edge whose removal disconnects the
graph and increases the number of connected components.
We calculate the fraction of bridges in a wide range of real-world networks and their randomized counterparts. We find that real networks typically have more bridges than their completely randomized counterparts, but very similar fraction of bridges as their degree-preserving randomizations. 
We define a new edge centrality measure, called bridgeness, to quantify the importance of a bridge in damaging a network. We find that certain real networks have very large average and variance of bridgeness compared to their degree-preserving randomizations and other real networks.
Finally, we offer an analytical framework to calculate the bridge fraction , the average and variance of bridgeness for uncorrelated random networks with arbitrary degree distributions. 
\end{abstract}


\maketitle


A \emph{bridge}, also known as \emph{cut-edge}, is an edge of a
graph whose removal disconnects the graph, i.e., increases the number
of connected components (see Fig. 1, red edges)~\cite{bollobas1998modern}. 
A dual concept is \emph{articulation point} (AP) or \emph{cut-vertex}, defined as a node in a graph whose removal disconnects the graph~\cite{behzad1972introduction,harary1969graph}. 
Both bridges and APs in a graph can be identified via a linear-time algorithm based on depth-first search~\cite{tarjan1972depth} (see Supplemental Material Sec.I for details) and represent natural vulnerabilities of real-world networks. Analysis of APs has recently provided us a new angle to systematically investigate the structure and function of many real-world networks~\cite{tian2016articulation}. This prompts us to ask if similar analysis can be applied to bridges. 

Note that bridge is similar but different from the notion of \emph{red bond} introduced in percolation theory to characterize substructures of percolation clusters on lattices~\cite{pike1981order}. To define a red bond, we consider the percolation cluster as a network of wires carrying electrical current and we impose a voltage drop between two nodes in the network. Then red bonds are those links that carry all current, whose removal stops the current. The definition of bridges does not require us to impose a voltage drop on the network. Instead, it just concerns the connectivity of the whole network. 

Despite that bridges play important roles in ensuring the network connectivity, the notion of bridge has never been systematically studied in complex networks. 
What is the typical number of bridges in a random graph with prescribed degree distribution? Are the bridges in a real network overpresented or underpresented? How to quantify the network vulnerability in terms of bridge attack? In this Letter, we systematically address those questions in both real networks and random graphs.

\begin{figure}[t]
\centering
\includegraphics[width=0.48\textwidth]{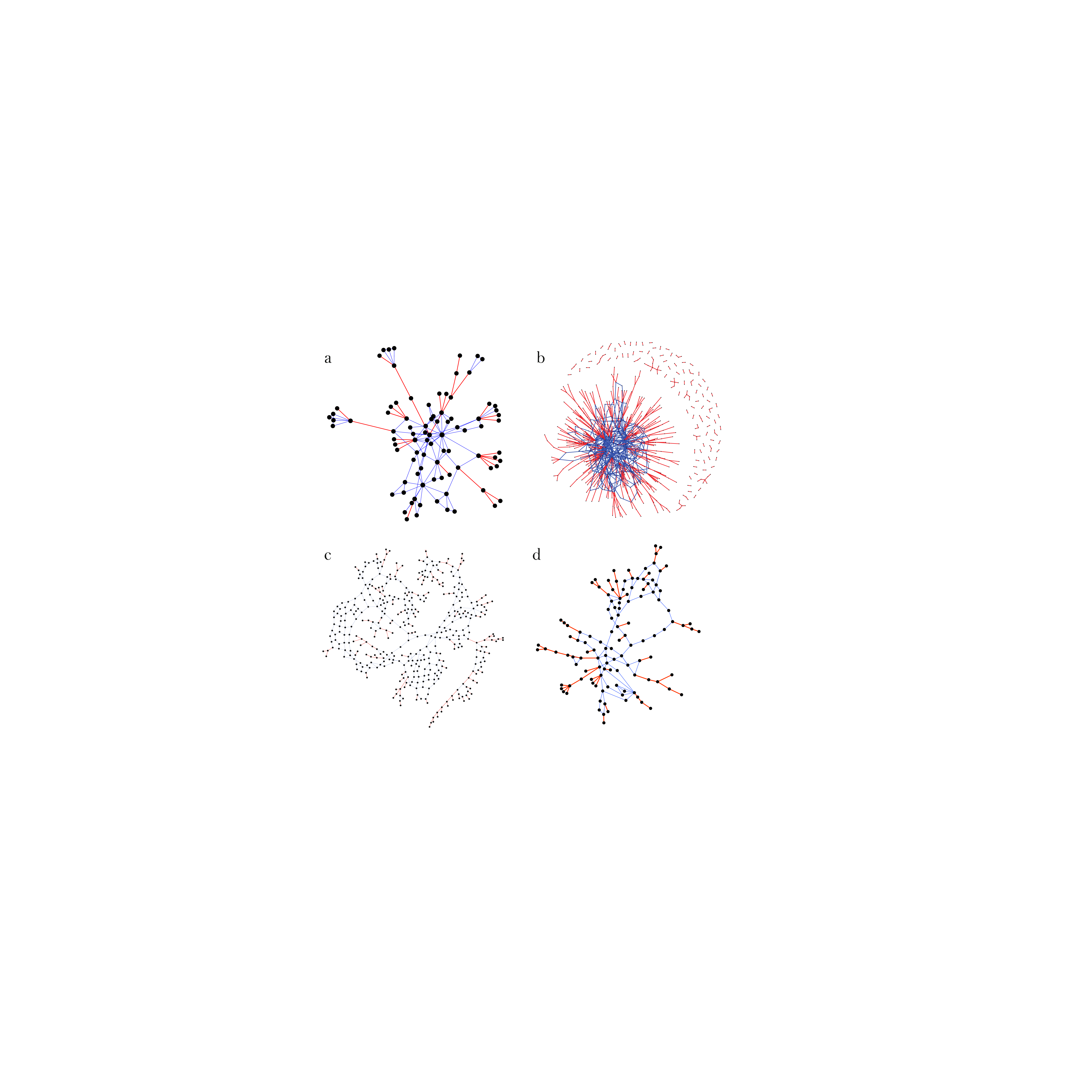} 
    \caption{ \label{fig:1}\textbf{Bridges in real-world networks.} Bridges (in red) are edges whose removal will increase the number of connected components in a graph. (\textbf{a}) Food web of \emph{Grassland}~\cite{dunne2002food}; (\textbf{b}) The protein-protein interaction network of \emph{C. elegans}~\cite{simonis2009empirically}; (\textbf{c}) A subgraph of the road network of California~\cite{leskovec2015snap}; (\textbf{d}) A subgraph of the power grid in three western states of US~\cite{watts1998collective}.} 
\end{figure}

 We first calculate the fraction of bridges ($f_\mm{b}:=L_\mm{b}/L$) in a wide range of real-world networks, from infrastructure networks to food webs, neuronal networks, protein-protein interaction (PPI) networks, gene regulatory networks, and social graphs. Detailed information of those networks can be found in Supplemental Material Sec. IV. Here $L_\mm{b}$ and $L$ denote the number of bridges and total links in a network, respectively. 
We find that many real networks have very small fraction of bridges, while a few of them (e.g., PPI networks) have very large fraction of bridges (Fig. 2a).  
To identify the topological characteristics that determine $f_\mm{b}$ in real networks, we compare $f_\mm{b}$ of a given network with that of its randomized counterpart. We first randomize each real network using a complete randomization procedure that turns the network into an Erd{\H{o}}s-R{\'e}nyi (ER) type of random graph with the number of nodes $N$ and links $L$ unchanged~\cite{erdos1960evolution}. 
We find that most of the completely randomized networks possess very different $f_\mm{b}$, compared to their corresponding real networks (Fig. 2a). This indicates that complete randomization eliminates the topological characteristics that determine $f_\mm{b}$. 
Moreover, we find that real networks typically display much higher $f_\mm{b}$ than their completely randomized counterparts (Fig. 2a). By contrast, when we apply a degree-preserving randomization, which rewires the edges among nodes while keeping the degree $k$ of each node unchanged, this procedure does not alter $f_\mm{b}$ significantly (Fig. 2b). In other words, the characteristics of a real network in terms of $f_\mm{b}$ is largely encoded in its degree distribution $P(k)$. 
\begin{figure}[t]
\centering
\includegraphics[width=0.48\textwidth]{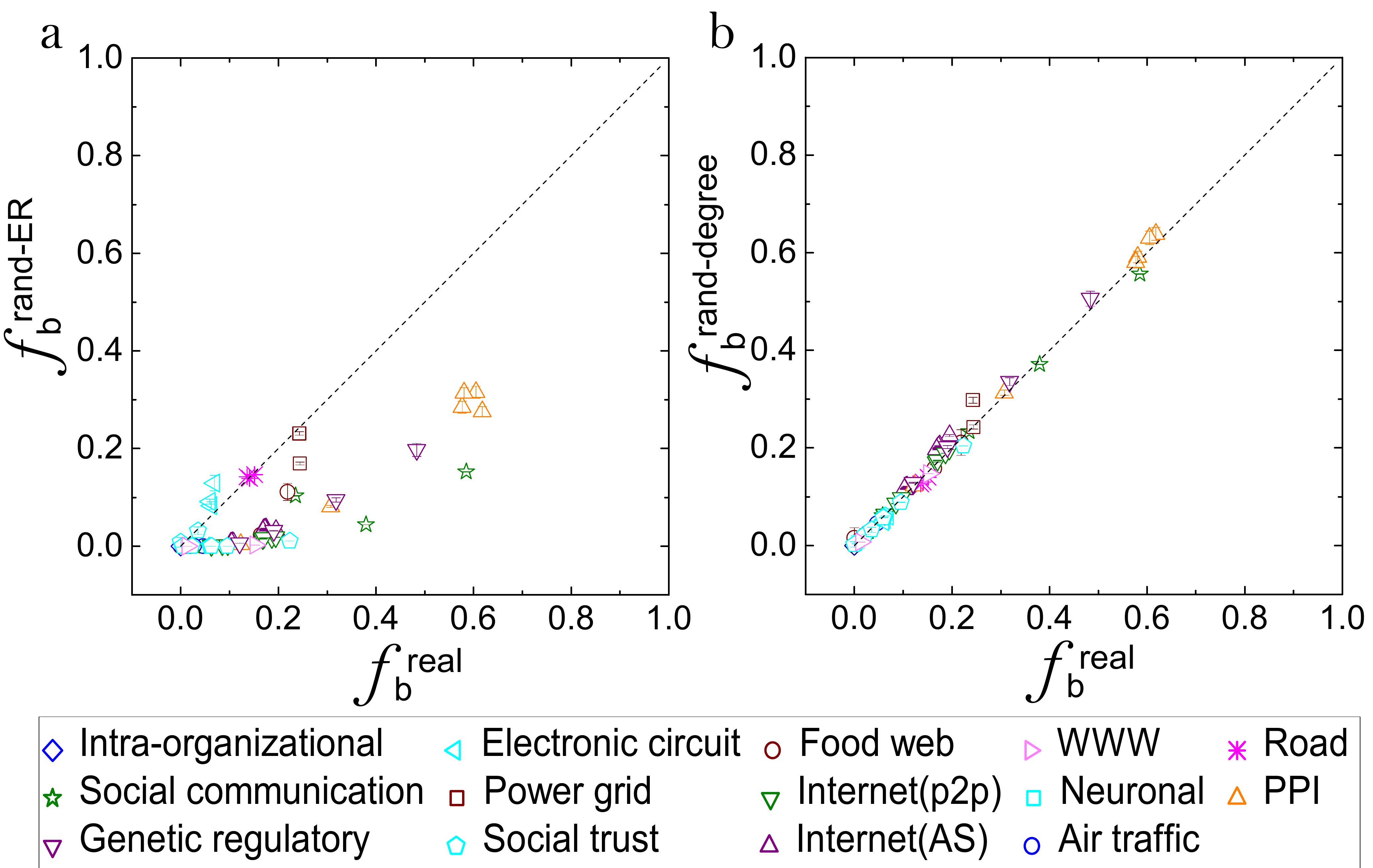}
\caption{\label{fig:2} \textbf{Fraction of bridges in real-world networks.} The dashed lines are $y=x$. The error bars represent the standard deviation, calculated from 100 randomizations. (\textbf{a}) Complete randomization of real networks. (\textbf{b}) Degree-preserving randomization.}
\end{figure}

In order to quantify the importance of an edge in damaging a network, we define an edge centrality measure $B$, called \emph{bridgeness}, for each edge in a graph as the number of nodes disconnected from the giant connected component (GCC)~\cite{bollobas2001cambridge} after the edge removal. By definition, if an edge is not a bridge or outside the GCC, it has zero bridgeness. Also, in the absence of GCC, all edges have zero bridgeness. We notice that bridgeness has been defined differently in the literature ~\cite{nepusz2008fuzzy,gong2011variability,jensen2015detecting,cheng2010bridgeness} (see Supplemental Material Sec.II). Here we define bridgeness based on the notion of bridge and we focus on the damage to the GCC, which is typically the main functional part of a network.

\begin{figure}[t]
\centering
\includegraphics[width=0.48\textwidth]{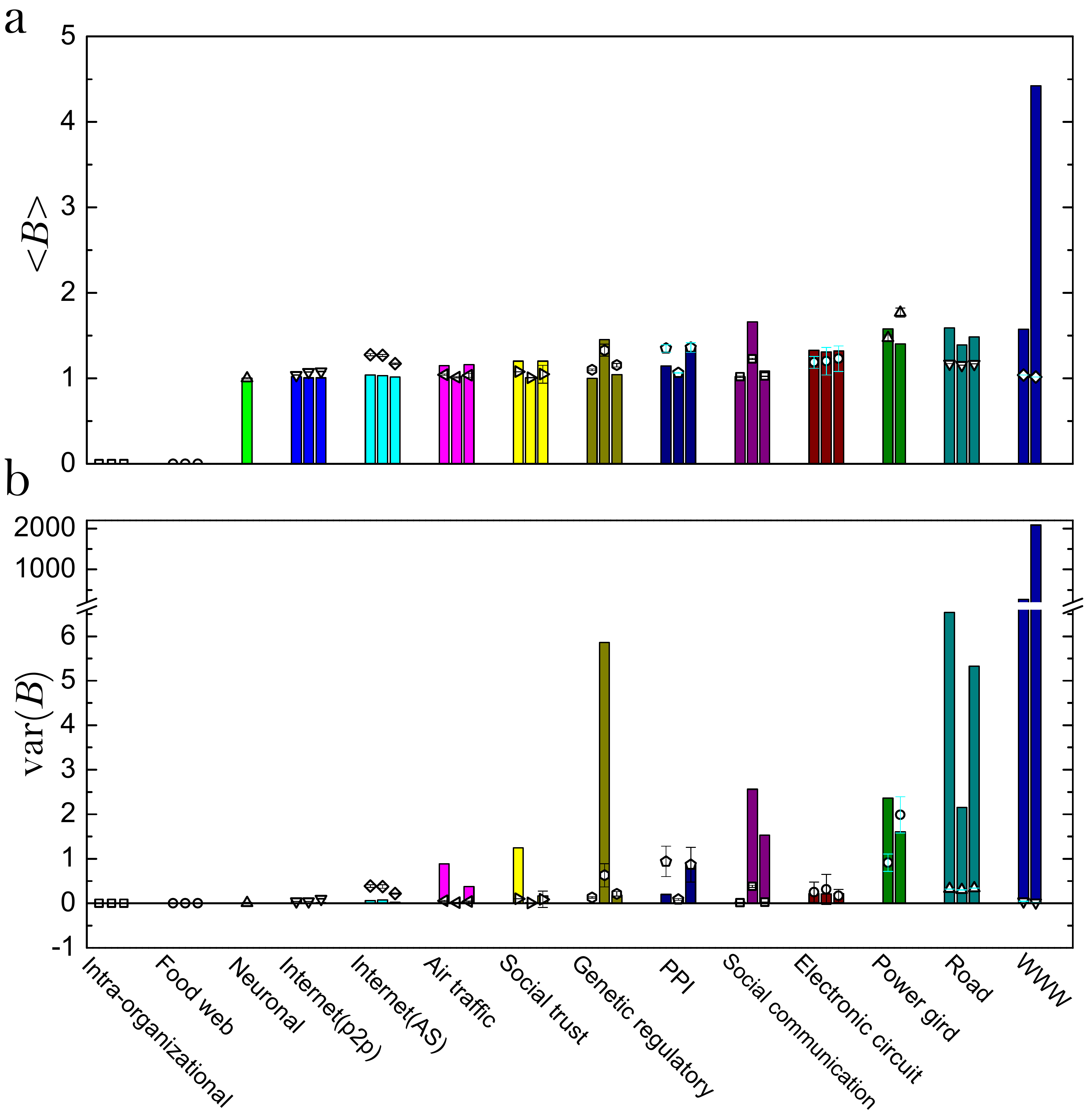}
\caption{\label{fig:4} \textbf{Average and variance of bridgeness in real-world networks.} The bars represent the values of real networks and empty symbols represent the values of their degree-preserving randomizations. The error bars represent the standard deviation, calculated from 10 randomizations. (\textbf{a}) Average bridgeness. (\textbf{b}) Variance of bridgeness.} 
\end{figure}
Bridgeness differentiates edges based on their structural importance. Consider all bridges that have non-trivial bridgeness, i.e., $B>0$. Denote their average and variance as $\langle B \rangle$ and $\mm{var}(B)$, respectively. We find that Word Wide Web (WWW) and road networks have much larger $\langle B \rangle$ than their randomized counterparts and other real networks (Fig. 3a). Moreover, those networks also have very large $\mm{var}(B)$ (Fig. 3b). The reason why road networks have very large $\langle B \rangle$ and $\mm{var}(B)$ is the presence of very long paths and the expense of constructing alternative paths. 
While for WWW, the reason is the presence of certain bridges that connect different large biconnected components in the GCC (see Supplemental Material Sec.IV). Here a biconnected component (BCC) is a connected subgraph where for any two nodes there are at least two paths connecting them that have no nodes in common other than these two nodes~\cite{RN8034}. (Note that by definition no bridges exist in a BCC.)

Since the bridge fractions in real networks are almost the same as
their degree-preserving randomized counterparts, the difference of
average bridgeness between real networks and their degree-preserving
randomizations indicates variations of vulnerability of those networks
in terms of bridge attack. Fig. 3a shows that certain types of
networks, such as air traffic, road networks, social graphs and WWW,
are more vulnerable, displaying much larger $\langle B \rangle$ than
their randomizations. By contrast,  the Internet at the autonomous
system (AS) level and the Internet peer-to-peer (p2p) file sharing
networks have smaller $\langle B \rangle$ than their randomized
counterparts, indicating that those networks are robust from the
bridge attack perspective. 

The results of real-world networks prompt us to analytically decipher bridge structure for large uncorrelated random networks with prescribed degree distributions. 
To begin with, we adopt the local tree approximation, which assumes the absence of finite loops in the thermodynamic limit (i.e., as the network size $N \to \infty$) and allows only infinite loops~\cite{tian2016articulation}. This approximation leads to three important properties: (1) all finite connected components (FCCs) are trees, and hence all edges inside them are bridges; (2) there exists only one giant connected component (GCC)~\cite{newman2001random}, only one BCC (which has no bridges), and the BCC is a subgraph of the GCC; (3) subgraphs inside the GCC but outside the BCC are trees and all edges in those subgraphs are bridges~\cite{cohen2000resilience,callaway2000network,newman2001random,dorogovtsev2008critical,mezard2009information,tian2016articulation}.

\begin{figure}[t]
\centering
\includegraphics[width=0.4\textwidth]{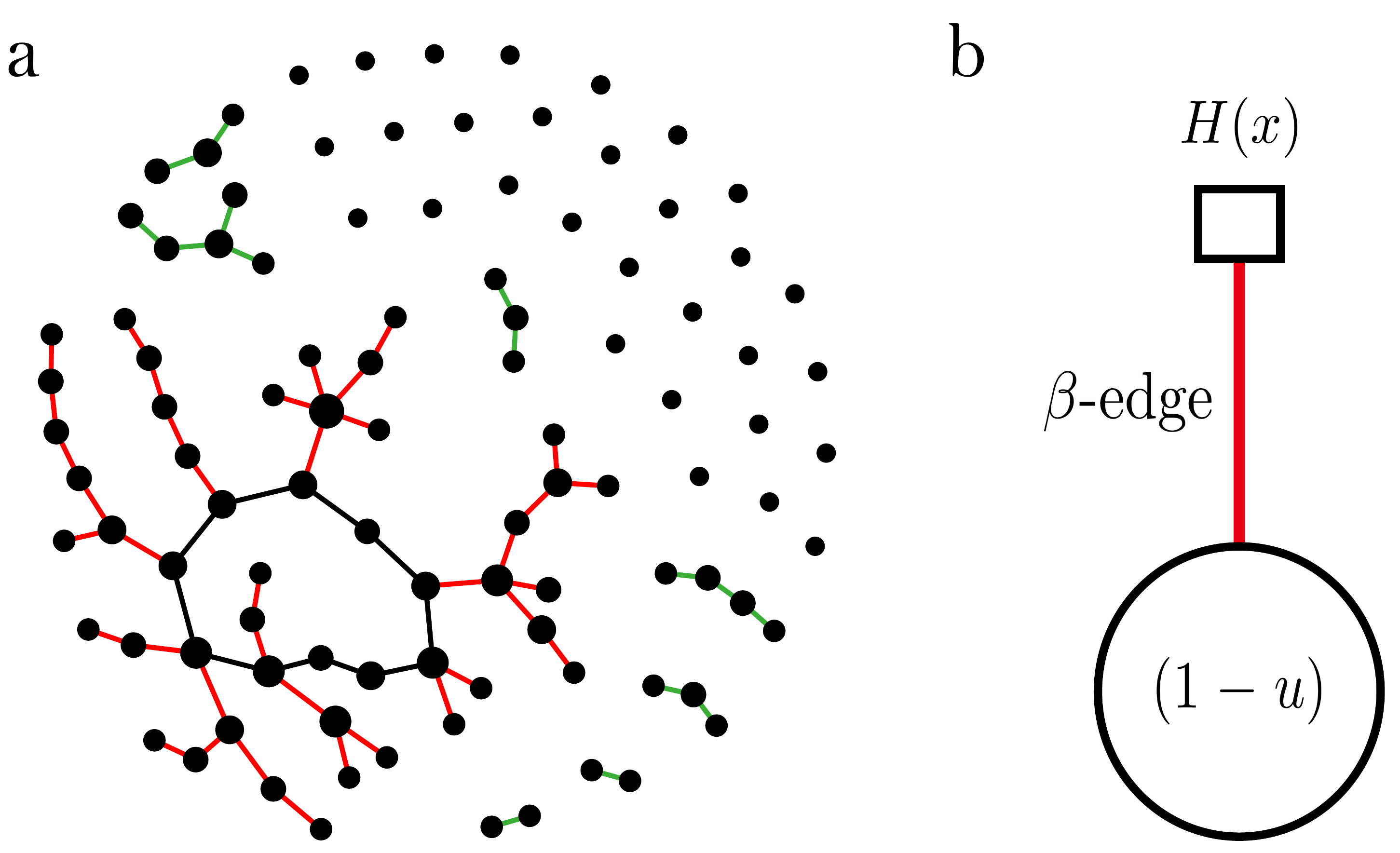}
\caption{\label{fig:5} \textbf{Demonstration of different types of edges.} (\textbf{a}) The green lines, red lines and black lines represent $\alpha$-edges, $\beta$-edges and $\gamma$-edges, respectively. (\textbf{b}) Neighborhood of a $\beta$-edge. The red line is a $\beta$-edge and the black square and ellipse represent an FCC and the GCC, respectively, after the removal of the $\beta$-edge.}
\end{figure}

Based on the above considerations, we categorize all the edges in a graph into three types (Fig. 4a):
  (i) $\alpha$-edge: edges in FCCs, which are bridges;
  (ii) $\beta$-edge: edges inside the GCC but outside the BCC, which are also bridges;
  (iii) $\gamma$-edge: edges inside the BCC, which are not bridges.
Hereafter we also use $\alpha$, $\beta$ or $\gamma$ to denote the probability that a randomly chosen edge is a $\alpha$-edge, $\beta$-edge, or $\gamma$-edge, respectively. By definitions, we have $\alpha + \beta + \gamma=1$, and $f_\mm{b}= \alpha + \beta$. Note that according to our definition of bridgeness, only $\beta$-edges have nontrivial bridgenesses, i.e., $B>0$.

The generating functions $G_0(x)=\sum_{k=0}^\infty P(k) x^k $ and $G_1(x)= \sum_{k=1}^\infty Q(k) x^{k-1} $ are very useful in calculating key quantities of random graphs, such as the mean component size and the size of GCC \cite{newman2001random}. Here $Q(k)=k P(k)/c$, and $c=\sum_{k=0}\infty k P(k)$ is the mean degree. To calculate $\alpha$, $\beta$ and $\gamma$, we introduce the generating function $H_1(x)$ for the size distribution of the components that are reached by choosing a random edge and following it to one of its ends. (Note that the notation $H_0(x)$ is reserved for the generating function of the size distribution of the components that a randomly chosen node sits in, see Supplemental Material Sec. III) ~\cite{newman2001random}. Note that we only include the FCCs in calculating $H_1(x)$, which means that the chosen edge must be a bridge, namely either $\alpha$- or $\beta$-edge. 

According to the local tree approximation, $H_1(x)$ satisfies the following self-consistency equation~\cite{newman2001random}:
$$ H_1(x)= \sum_{k=1}^\infty xQ(k)[H_1(x)]^{k-1}.\eqno {(1)}$$
 Equation (1) implies that following a bridge, the excess edges of its end to finite subcomponents should also be bridges. We can rewrite Eq. (1) using the generation function of $Q(k)$, i.e., $$H_1(x)=xG_1(H_1(x)).\eqno {(2)}$$

Define $u:=H_1(1)$, which represents the probability that following a randomly chosen edge to one of its end nodes, the node belongs to an FCC after removing this edge.
Then the probability that a randomly chosen edge is an $\alpha$-edge or belongs to an FCC is  simply $\alpha=u^2$.
    For a $\beta$-edge, one of its end nodes belongs to an FCC and the other one belongs to the GCC after removing this edge. Hence we have  $\beta=2u(1-u).$
   For a $\gamma$-edge, both of its end nodes belong to the GCC after its removal, and hence $\gamma=(1-u)^2.$
Note that the normalization condition $\alpha+\beta+\gamma=1$ is naturally satisfied.
The fraction of bridges is simply given by 
$$ f_\mm{b}=\alpha+\beta=1-(1-u)^2.\eqno {(3)}$$

In Fig. 5a, we show the bridge fraction $f_\mm{b}$ calculated from Eqs.(2) and (3), the relative size of BCC ($s_\mm{BCC}$)~\cite{RN8034}, and the relative size of GCC ($s_\mm{GCC}$)~\cite{newman2001random} as functions of mean degree $c$ in ER random graphs with Poisson degree distribution $P (k ) = e^{-c} c^k / k! $~\cite{erdos1960evolution}. We find that
before the GCC and BCC emerge at the percolation threshold $c^*=1$, all components are FCCs and all edges are $\alpha$-edges, rendering $f_\mm{b}=1$. After the emergence of the GCC and BCC at $c^* = 1$~\cite{dorogovtsev2008critical}, $f_b$ begins to deviate from $1$, and the fraction of $\beta$-edges displays a non-monotonic behavior (because the difference between $s_\mm{GCC}$ and $s_\mm{BCC}$ increases first and then decreases). We also calculate $f_\mm{b}$ for scale-free (SF) networks with power-law degree distribution $P (k ) \sim k^{-\lambda}$ generated by the static model~\cite{goh2001universal,catanzaro2005analytic,lee2006intrinsic}. For SF networks, the smaller the degree exponent $\lambda$, the smaller the percolation threshold $c^*$~\cite{dorogovtsev2008critical}, rendering $f_\mm{b}$ deviate from 1 at smaller $c^*$ (Fig. 5b).

\begin{figure}[t]
\centering
\includegraphics[width=0.48\textwidth]{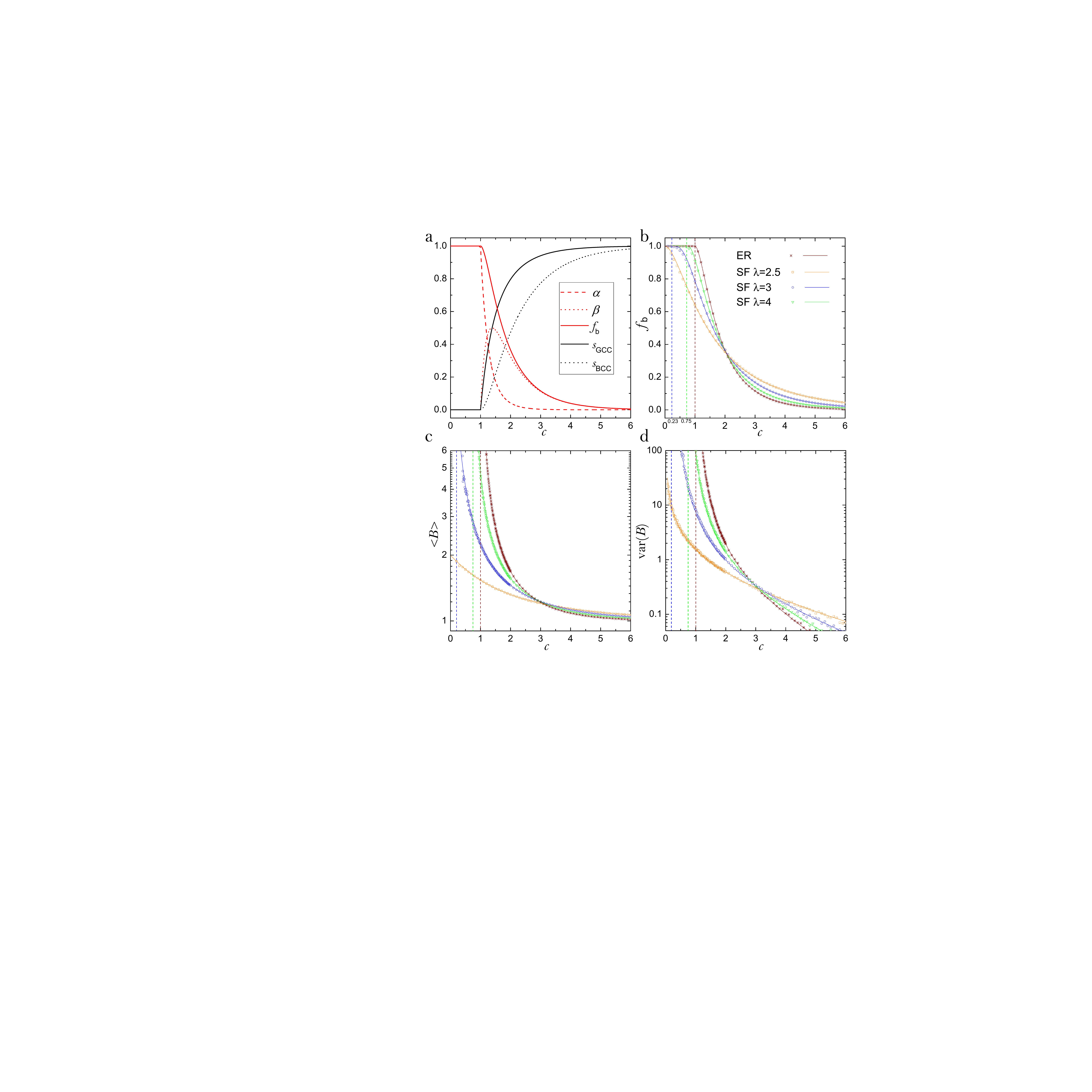}
\caption{\label{fig:3} \textbf{Bridge and bridgeness in ER and SF random networks.} The size of all these random network is $10^6$. Curves in (\textbf{a})-(\textbf{d}) are analytical predication about bridges, and symbols are simulation results. (\textbf{a}) The behavior of the bridge fraction $f_\mm{b}$, relative sizes of the BCC and GCC (denoted as $s_\mm{BCC}$, $s_\mm{GCC}$, respectively) in ER random graphs. (\textbf{b}) Bridge fraction in different random networks with dashed vertical lines representing the corresponding percolation threshold $c^*$ where the GCC and BCC emerge. Note that $c^*_\mm{SF}(\lambda=2.5) = 0$, $c^*_\mm{SF}(\lambda=3.0) \approx 0.23$, $c^*_\mm{SF}(\lambda=4.0) \approx 0.75$, and $c^*_\mm{ER}=1.0$. (\textbf{c}) Average bridgeness in random networks. (\textbf{d}) The variance of bridgeness in random networks. For detailed calculation and more distributions see Supplemental Material Sec. III.}
\end{figure}

Besides $f_\mm{b}$, we can also calculate the bridgeness distribution $P(B)$ from $H_1(x)$. For nontrivial bridgeness ($B>0$) we only consider the bridges in the GCC. In other words, we calculate $P(B)$ for $\beta$-edges in random graphs. Define the generating function of $P(B)$ as 
$$F(x) = \sum_{B=1}^\infty P(B) x^B, \eqno {(4)}$$
which leads to $P(B)=\frac{1}{B!} \frac{\mathrm{d}^BF(x)}{\mathrm{d}x^B}|_{x=0}.$
Since one end node of a $\beta$-edge locates in the GCC after this edge is removed (Fig. 4b), we have:
$$F(x)=\frac{2(1-u)H_1(x)}{\beta}=\frac{H_1(x)}{u}, \eqno {(5)} $$
where the numerator represents the generating function for the bridgeness distribution of a randomly chosen $\beta$-edge, and the denominator originates from the fact that we focus on $\beta$-edges.
The moments of $P(B)$ are then given by:
$$\langle B^k \rangle=\sum_{B=1}^\infty B^kP(B)=\left [ \left (x\frac{\mathrm{d}}{\mathrm{d}x}\right )^kF(x)\right ]_{x=1}.\eqno {(6)}$$ 

We calculate the average bridgeness $\langle B \rangle$ and the variance of bridgeness $\mm{var}(B) (:= \langle B^2 \rangle - \langle B \rangle^2)$ in ER and SF random networks (Fig. 5c-d).
We find that for both ER and SF networks, $\langle B \rangle$ and $\mm{var}(B)$ monotonically decrease as $c$ increases. Note that $\langle B \rangle$ and $\mm{var}(B)$ of SF networks are typically lower than those of ER networks for small $c$, and higher for large $c$.
This is because SF networks tend to first form densely connected components of hub nodes and then slowly stretch out. This means that they form the BCC earlier but extending bridges while ER networks absorbs bridges quickly. The divergent behavior of bridgeness around the percolation threshold $c^*$ is due to the emergence of the GCC, which initially is tree-like and therefore contains bridges with a huge range of bridgeness.

In conclusion, we systematically investigate the bridge structure in complex networks. We demonstrate bridges in real-world networks, calculate the fraction of bridges in different networks, and define a new edge centrality measure, called bridgeness, to quantify the importance of bridges in damaging a network. Finally we analytically calculate bridge structure in random graphs with prescribed degree distributions. The presented results help us understand the complex architecture of real-world networks and may shed lights on the design of more robust networks against bridge attack.

\emph{Acknowledgements:} We thank Wei Chen for valuable discussions. This work is supported by the John Templeton
Foundation (Award number 51977), National Natural Science Foundation of China (Grant No. 11505095), Research Fund for the Doctoral Program of Higher Education of China (Grant No. 20133218120033), and the Fundamental Research Funds for the Central Universities of China (Grants Nos. NS2014072 and NZ2015110).
\emph{Author Contributions:} Y.-Y.L conceived and designed the
project. A.-K.W and L.T. did the analytical calculations. A.-K.W. did the
numerical simulations and analyzed the empirical data. All authors
analyzed the results. A.-K.W. and Y.-Y.L. wrote the manuscript. L.T. edited
the manuscript.


%
\putbib
\end{bibunit}

\begin{bibunit}
\clearpage
\appendix
\section{Supplemental Material}
\section{Algorithm for identifying bridges and Calculating Bridgeness}
The algorithm for identifying bridges in a network is based on depth-first search (DFS), which has linear time complexity \cite{tarjan1972depth}. Randomly choosing a node from the network, we start DFS and track two indices for each node $i$: its DFS visited time stamp (DFS[$i$]) and the lowest DFS reachable ancestor (low[$i$]). 
$\text{DFS}[i]$ is defined as the number of other visited nodes till the current one in DFS. And $\text{low}[i]$ represents the lowest $\text{DFS}[j]$ of an previously visited node $j$ that can be reached again by current node $i$ in the later DFS. Note that, for two successively visited nodes $i$ and $j$ in the DFS, the index low[$i$] is updated by min(low[$i$], low[$j$]) after $j$ is visited.

Note that $\text{low}[i]$ marks the node's topological position in the network. 
For two nodes $i$ and $j$ in the same biconnected component (BCC), low[$i$]=low[$j$]. For nodes in tree structure, low[$i$]=DFS[$i$], which is different for each node. A bridge between two nearest neighboring nodes ($i$ and $j$) is identified whenever the later visited node, say node $i$, has larger low[$i$] than that of the previously visited node $j$. 

\begin{figure}
\centering
\includegraphics[width=0.48\textwidth]{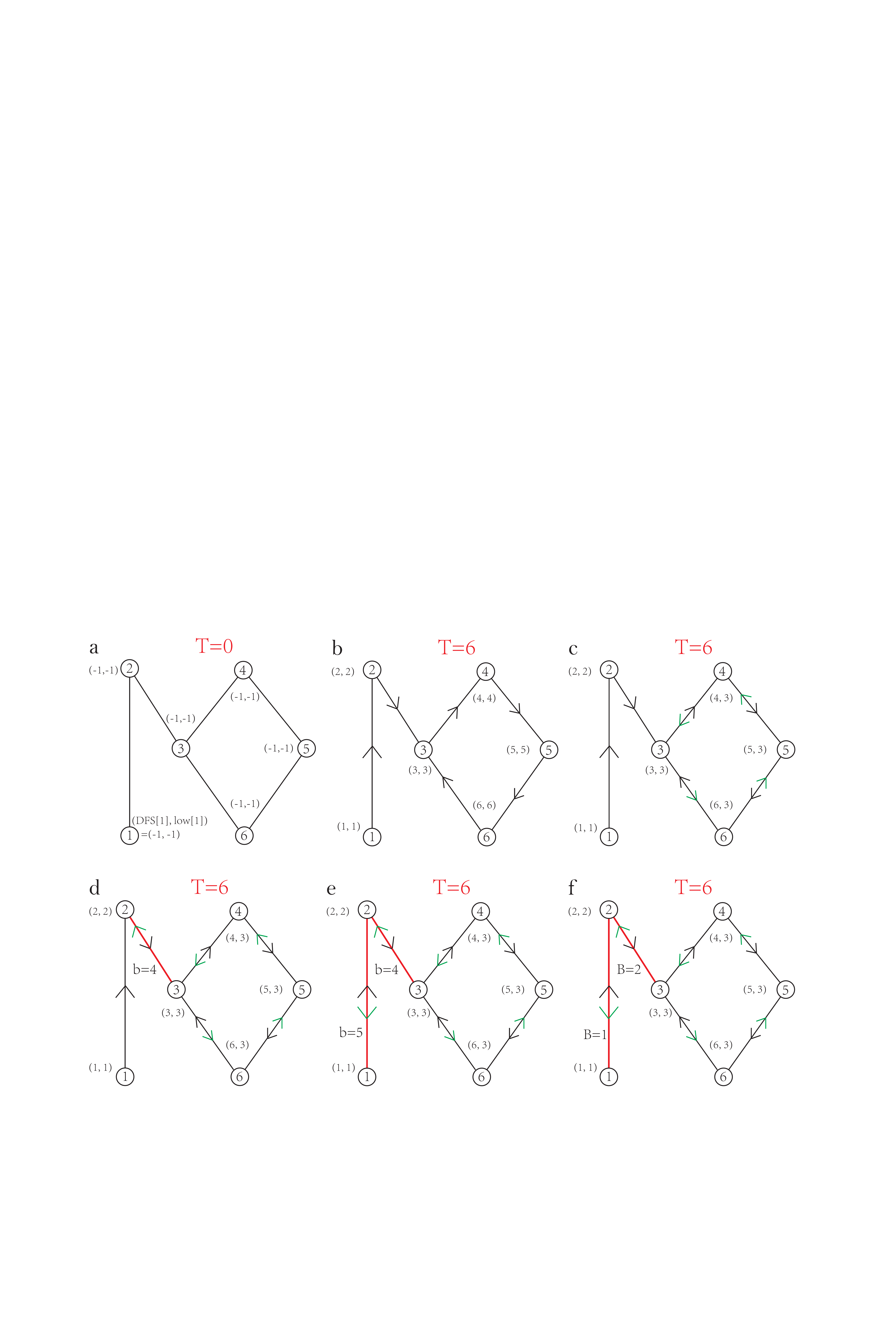}
\caption{\label{fig:0} \textbf{Using depth-first search to identify bridges and calculate bridgeness.} The labels in nodes and arrows on edges represent the sequence of DFS, red edges are bridges, and for each node its two indices are presented as a coordinate (DFS[i], low[i]). (\textbf{a}) Initial state of indices. (\textbf{b}) The moment when DFS just finishes visiting all the nodes in the connected component. The size $S$ of this connected component is given by the current time step $T$.  (\textbf{c-e}) Updates of indices when the search goes back. (\textbf{f}) Checking through all nodes in the component and let bridgeness be the size of the smaller part separated from that component.}
\end{figure}

To calculate the size ($b$) of the subgraph that will be cut from the network due to the removal of a bridge, we can simply use current time step ($T$), i.e., the number of visited nodes, to subtract the DFS visited time stamp of the end of the bridge (which is inside the BCC), and plus one. For instance, in Fig. \ref{fig:0}(c), $b=T-\text{DFS}[3]+1=6-3+1=4$.
Naturally each bridge has two components to be cut from the network. And we define bridgeness to be the smaller size of the two components. Thus, to calculate the bridgeness, we need to go through the giant connected component (GCC) again, and $B$is calculated as $\text{min} \{ b, S-b \}$, where $S$ represents the size of the whole connected component, see Fig. \ref{fig:0}(e).

To summarize, we first conduct DFS in each connected component of a graph to identify bridges with one of the separating parts ($b$) after their removal and get the size of each connected component. Then we go through the GCC again to get the bridgeness ($B$) of each $\beta$-edge.

\section{Previous definitions of bridgeness}
The notion of bridgeness has been introduced in the literature with various different definitions. But none of them is based on the notion of bridges. Some of them are actually node-based. 

\subsection{A local index on edge significance in maintaining global connectivity}
\begin{figure}
\centering
\includegraphics[width=0.48\textwidth]{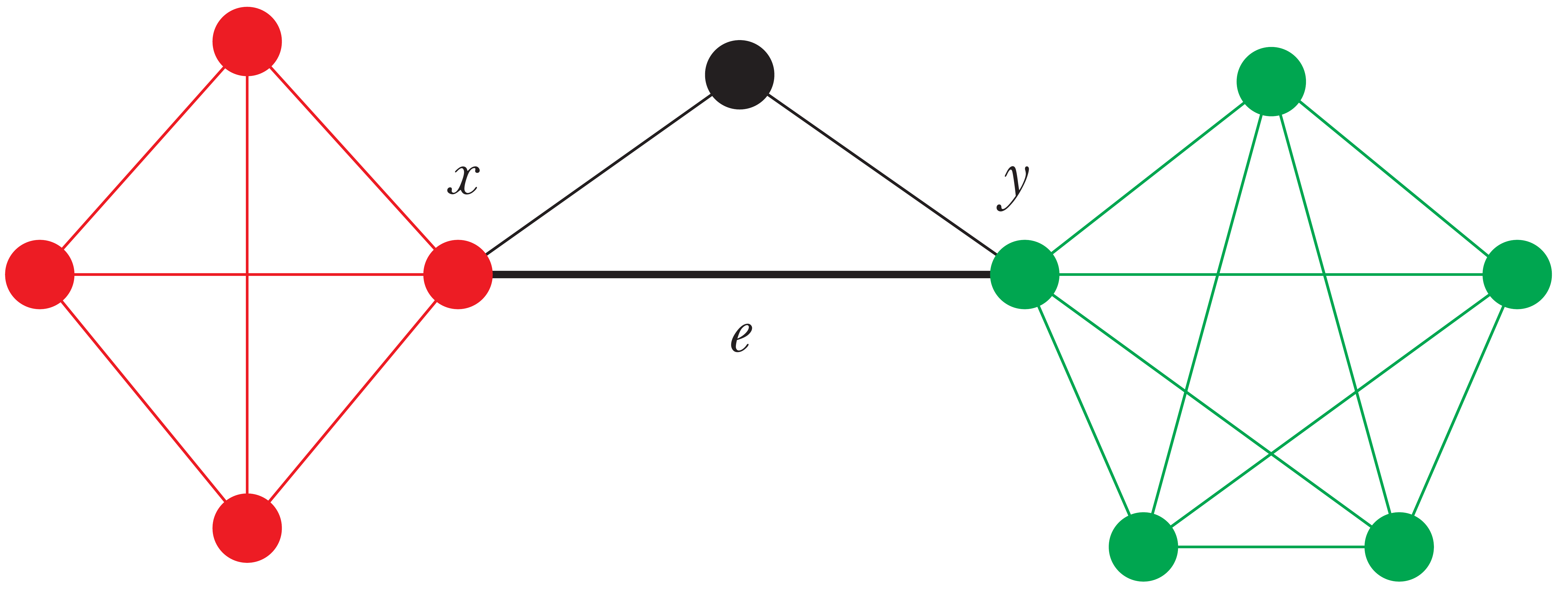}
\caption{\label{fig:0.5} \textbf{An example for the local bridgeness.} In this example, $S_x=4$, $S_y=5$, $S_e=3$ and $B_e=1.49$.}
\end{figure}

In \cite{cheng2010bridgeness}, the bridgeness of an edge is defined to be a local index quantifying the edge importance in maintaining the network connectivity: 
 \begin{equation}
	 B_e=\frac{\sqrt{S_xS_y}}{S_e},
\end{equation}
where $x$ and $y$ are the two endpoints of the edge $e$ and $S_x$, $S_y$, $S_e$ are the clique sizes of nodes $x$, $y$ and the edge $e$, respectively. A clique of size $k$ is a fully connected subgraph with $k$ nodes \cite{xiao2007empirical} and the clique size of a node $x$ or an edge $e$ is defined as the size of maximum clique that contains this node or edge \cite{shen2009detect,shen2009quantifying}. See Fig. \ref{fig:0.5} for a small example. 

\subsection{Global bridges in networks}
\begin{figure}
\centering
\includegraphics[width=0.48\textwidth]{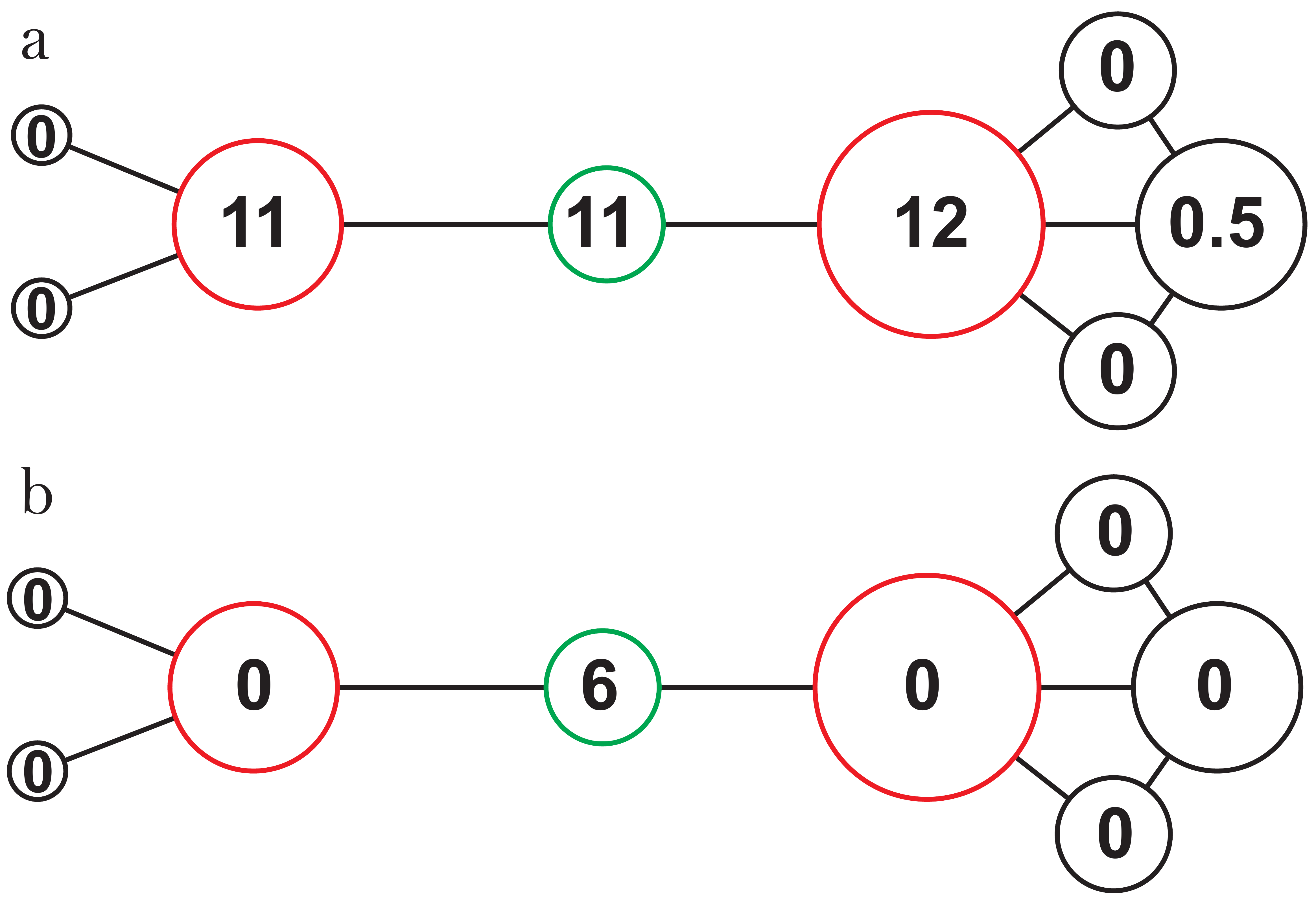}
\caption{\label{fig:0.6} \textbf{Betweenness centrality and bridgeness centrality.} The sizes of nodes are proportional to their degrees and colored nodes are articulation points, where the green one is the global center and red ones are local centers. Numbers in nodes are their BC or BRI measures. (\textbf{a}) Betweenness centrality in a graph. (\textbf{b}) Bridgeness centrality in the same graph. 
Betweenness cannot differentiate the global center (the green node) from APs  (red nodes), as it gives slighter higher score to high-degree nodes, which are local centers. In contrast, bridgeness  centrality effectively scores the node that plays an important role in global connectivity (the green node).}
\end{figure}

A node-based bridgeness, called bridgeness centrality (BRI), is derived from the node's betweenness centrality (BC) \cite{jensen2015detecting}. Consider the betweenness centrality of a node $j$ \cite{freeman1977set,freeman1979centrality}:
 \begin{equation}
	 BC(j)=\sum_{i \neq j \neq k} \frac{\sigma_{ik}(j)}{\sigma_{ik}},
\end{equation}
where $i$, $j$, $k$ are nodes; $\sigma_{ik}$ represents the number of shortest paths between $i$ and $k$ while $\sigma_{ik}(j)$ is the number of such paths running through $j$.
The bridgeness centrality of node $j$ is defined as the non-local part of its betweenness centrality: 
 \begin{equation}
	 BRI(j)=\sum_{i \notin N_G(j) \wedge k \notin N_G(j)} \frac{\sigma_{ik}(j)}{\sigma_{ik}},
\end{equation}
where $N_G(j)$ are neighbor nodes of $j$. Examples are shown in Fig. \ref{fig:0.6}.

\subsection{Nodal bridgeness in communities with overlap }
Nodal bridgeness can also be defined as a generalization of articulation point to solve the community detection problem. The number of communities $M$ in a graph can either be given in advance or by some community detection algorithm \cite{nepusz2008fuzzy} and the partition of nodes is represented by the partition matrix $\textbf{U}=[u_{ik}]$, where $u_{ik}$ measures the relationship between the node $v_k$ and community $i$, which is determined by the complicated partition based on vertex similarities \cite{nepusz2008fuzzy}. 
This nodal bridgeness measures the extent, to which a given node is shared among different communities \cite{nepusz2008fuzzy}. 
If a node belongs only to one community, it has zero bridgeness while a node shared by all communities has bridgeness one. This bridgeness is defined on a vertex $v_i$ as the distance of its membership vector $\textbf{u}_i=[u_{1i},u_{2i},...,u_{Mi}]$ from the reference vector $[1/M,1/M,...1/M]$ in the Euclidean vector norm, inverted and normalized to $[0,1]$ as \cite{nepusz2008fuzzy}:
 \begin{equation}
	 b_i=1-\sqrt{\frac{M}{M-1}\sum_{j=1}^{M} \left ( u_{ji}-\frac{1}{M} \right )^2}.
\end{equation}

\section{Bridges in random graphs with specific degree distributions}

In this section, we derive the equations in analytically calculating the first and second moments of the bridgeness distribution, as well as the relative size of GCC and BCC, for uncorrelated random graphs with prescribed degree distributions.

According to the definitions of $G_1(x)=\sum_{k=1}^\infty Q(k)x^{k-1}$, $H_1(x)= \sum_{k=1}^\infty xQ(k)[H_1(x)]^{k-1}$, $Q(k)=kP (k) /c$ and the self-consistency equation $H_1(x)=xG_1(H_1(x))$, we calculate $H_1'(X)$ and $H_1''(x)$ as follows:  
 \begin{equation}\label{h'}
 \begin{split}
	 H_1'(x)=G_1(H_1(x))+xG_1'(H_1(x))H_1'(x)\\ \Rightarrow H_1'(x)=\frac{G_1(H_1(x))}{1-xG_1'(H_1(x))},
 \end{split}
\end{equation}
 and 
\begin{equation}\label{h''}
\begin{split}
	 & H_1''(x)= \{ G_1'(H_1(x))H_1'(x)[1-xG_1'(H_1(x))]+\\
     & \qquad \qquad [G_1'(H_1(x))+xG_1''(H_1(x))H_1'(x)]G_1(H_1(x))  \} / \\ 
     &\qquad \qquad [1-xG_1'(H_1(x))]^2\\ 
     & \qquad \quad =\{ G_1'(H_1(x))G_1(H_1(x))+\\
     & \qquad \qquad [G_1'(H_1(x))+xG_1''(H_1(x))H_1'(x)]G_1(H_1(x)) \} / \\
     &   \qquad \qquad  [1-xG_1'(H_1(x))]^2.
\end{split}
\end{equation}
Therefore we have:
 \begin{equation}\label{aveB}
	\langle B \rangle= F'(1)=\frac{H_1'(1)}{u}=\frac{H_1'(1)}{H_1(1)},
\end{equation}
with $u=H_1(1)$ is the probability that following a randomly chosen edge to
one of its end nodes, the node belongs to an FCC after removing this edge. And
 \begin{equation}\label{aveB2}
 \begin{split}
	 & \langle B^2 \rangle=\left [ \left (x\frac{\mathrm{d}}{\mathrm{d}x}\right )^2F(x)\right ]_{x=1}=F'(1)+F''(1)\\
     & \qquad  =\frac{H_1'(1)+H_1''(1)}{u}.
 \end{split}
\end{equation}
Consequently, the variance of bridgeness is
 \begin{equation}\label{varB}
	 \text{var}(B) := \langle B^2 \rangle - \langle B \rangle^2=\frac{H_1'(1)[u-H_1'(1)]+uH_1''(1)}{u^2}.
\end{equation}

Note that $G_0(x)=\sum_{k=0}^\infty P(k)x^k$ is the generating function of the node degree distribution $P(k)$ and $H_0(x)=\sum_{k=0}^\infty xP(k)H_1^k(x)=xG_0(H_1(x))$ is the generating function for the size distribution of components that a randomly chosen node sits in. For the calculation of the relative size of GCC, we let $s_\text{FCC}$ be the fraction of vertices in the graph that do not belong to the giant component. Hence we have
 \begin{equation}
	s_\text{FCC}=H_0(1)=G_0(u).
\end{equation}
Then the relative size of the GCC is given by
 \begin{equation}\label{GCC}
	s_\text{GCC}=1-s_\text{FCC}= 1-G_0(u).
\end{equation}

For the calculation of the relative size of BCC, it can be derived as \cite{newman100bicomponents}
 \begin{equation}\label{GCC}
  \begin{split}
	& s_\text{BCC}=1-\sum_{k=1}^\infty P(k)u^k-\sum_{k=1}^\infty kP(k)(1-u)u^{k-1}\\
    &  \qquad =1-G_0(u)-(1-u)G_0'(u),
 \end{split}
\end{equation}
where $\sum_{k=1}^\infty P(k)u^k+\sum_{k=1}^\infty kP(k)(1-u)u^{k-1}$ means that if a vertex is outside the BCC, its surroundings should have at most one element that is not $u$.

Here we propose a new method to calculate $s_\text{BCC}$, which relies on the result of  $s_\text{GCC}$. Consider the $\beta$-edges, which are inside the GCC but outside of the BCC. Note that each $\beta$-edge can be assigned to one node that is inside the GCC but outside the BCC. Hence $s_\text{BCC}$ can be calculated as:
 \begin{equation}\label{BCC}
	s_\text{BCC}=s_\text{GCC} - \beta c/2 =s_\text{GCC} - u(1-u)c,
\end{equation}
where $\beta c/2 = u(1-u)c$ represents the fraction of $\beta$-edges normalized by total number of nodes. Note that the above two equations are equivalent, because $G_0'(u)=cG_1(u)=cu$.

\subsection{Poisson-distributed graphs}
 The degree distribution $P(k)$ for Erd{\H{o}}s-R{\'e}nyi random graphs follows Poisson distribution \cite{erdos1960evolution}: 
 \begin{equation}
	 P(k)=\frac{e^{-c}c^k}{k!},
\end{equation}
with $c$ is the mean degree.
 Then the generating functions are:
 \begin{equation}
	 G_0(x,c)=e^{c(x-1)},
\end{equation}
\begin{equation}
	 G_1(x,c)=e^{c(x-1)},
\end{equation}
with derivatives:  
\begin{equation}
	G_1'(x)=ce^{c(x-1)},
\end{equation}
\begin{equation}
	G_1''(x)=c^2e^{c(x-1)}.
\end{equation} 
With
\begin{equation}
	 u=H_1(1)=G_1(H_1(1))=e^{c(u-1)},
\end{equation}
and 
\begin{equation}
	 H_1'(1)=\frac{G_1(u)}{1-G_1'(u)}=\frac{u}{1-cu},
\end{equation}
we have 
\begin{equation}
	 f_{\text{b}}=[1-(1-u)^2],
\end{equation}
and 
\begin{equation}
	 \langle B \rangle=\frac{1}{1-cu}.
\end{equation}
 Substituting Eq. (S16-22) into Eq. (S6-9), we can get 
 \begin{equation}
	 \text{var}(B)=\frac{cu}{(1-cu)^3}.
\end{equation}
Besides, by Eq. (S11-13, 19), we also have 
 \begin{equation}
	 s_\text{GCC}=1-e^{c(u-1)},
\end{equation}
and 
 \begin{equation}
	s_\text{BCC}=1-e^{c(u-1)}  - u(1-u)c.
\end{equation}
Results are shown in main text Fig. 5.

\subsection{Exponentially distributed graphs}
The degree distribution for exponentially distributed graphs is \cite{liu2012core,newman2001random} :
\begin{equation}
	P(k)=(1-e^{-1/\kappa})e^{-k/\kappa},
\end{equation}
and the mean degree is 
\begin{equation}
	c=\frac{e^{-1/\kappa}}{1-e^{-1/\kappa}}.
\end{equation}
The generating functions are 
\begin{equation}
	G_0(x)=\frac{1-e^{-1/\kappa}}{1-xe^{-1/\kappa}},
\end{equation}
\begin{equation}
	G_1(x)=\left [\frac{1-e^{-1/\kappa}}{1-xe^{-1/\kappa}} \right ]^2,
\end{equation}
with derivatives  
\begin{equation}
	G_1'(x)=\frac{2e^{-1/\kappa}(1-e^{-1/\kappa})^2}{(1-xe^{-1/\kappa})^3},
\end{equation}
\begin{equation}
	G_1''(x)=\frac{6e^{-2/\kappa}(1-e^{-1/\kappa})^2}{(1-xe^{-1/\kappa})^4}.
\end{equation}

Inserting Eq. (S26-31) into Eq. (S5-13), we can get $f_\text{b}$, $s_\text{GCC}$, $s_\text{BCC}$, $\langle B \rangle$ and $\text{var}(B)$. Results of these quantities can be found in Fig. \ref{fig:1},\ref{fig:2}.

\subsection{Purely power-law distributed graphs}
The degree distribution for purely power-law distributed graphs is \cite{liu2012core,newman2001random}:
\begin{equation}
	P(k)=\frac{k^{-\lambda}}{\zeta(\lambda)} \qquad \text{for} \qquad k\ge 1,
\end{equation}
where $\zeta(\lambda)=\sum_{k=1}^{\infty}k^{-\lambda}$ is the Riemann Zeta function. Note that $P(k)$ can be normalized only for $\lambda \ge 2$. It is obvious that the mean degree is larger than $1$ in this situation and larger $\lambda$ leads to smaller mean degree.

The generating functions are 
\begin{equation}
	G_0(x)=\frac{\text{Li}_{\lambda}(x)}{\zeta(\lambda)},
\end{equation}
\begin{equation}
	G_1(x)=\frac{\text{Li}_{\lambda-1}(x)}{x\zeta(\lambda-1)},
\end{equation}
where $\text{Li}_{n}(x)=\sum_{k=1}^{\infty} x^k/k^{n}$ is the $n$th polylogarithm of $x$, whose derivative is $\frac{\mathrm{d}\text{Li}_{n}(x)}{\mathrm{d}x}= \frac{\text{Li}_{\lambda-1}(x)}{x}$.
The derivatives of the generating functions are
\begin{equation}
	G_1'(x)=\frac{\text{Li}_{\lambda-2}(x)-\text{Li}_{\lambda-1}(x)}{x^2\zeta(\lambda-1)},
\end{equation}
\begin{equation}
	G_1''(x)=\frac{\text{Li}_{\lambda-3}(x)-3\text{Li}_{\lambda-2}(x)+2\text{Li}_{\lambda-1}(x)}{x^3\zeta(\lambda-1)}.
\end{equation}

Inserting Eq. (S32-36) into Eq. (S5-13), we can get $f_\text{b}$, $s_\text{GCC}$, $s_\text{BCC}$, $\langle B \rangle$ and $\text{var}(B)$. Results are shown in Fig. \ref{fig:1},\ref{fig:2}.

\subsection{Power-law distribution with exponential cutoff}

The degree distribution for a purely power-law distribution with exponent $\lambda$ and exponential cutoff is \cite{liu2012core,newman2001random}:
\begin{equation}
	P(k)=\frac{k^{-\lambda}e^{-k/\kappa}}{\text{Li}_{\lambda}(e^{-1/\kappa})} \qquad \text{for} \qquad k\ge 1.
\end{equation}
This distribution can be normalized for any $\lambda$. 
 
The generating functions are 
\begin{equation}
	G_0(x)=\frac{\text{Li}_{\lambda}(xe^{-1/\kappa})}{\text{Li}_{\lambda}(e^{-1/\kappa})}.
\end{equation}
\begin{equation}
	G_1(x)=\frac{\text{Li}_{\lambda-1}(xe^{-1/\kappa})}{x\text{Li}_{\lambda-1}(e^{-1/\kappa})},
\end{equation}
\begin{equation}
	G_1'(x)=\frac{\text{Li}_{\lambda-2}(xe^{-1/\kappa})-\text{Li}_{\lambda-1}(xe^{-1/\kappa})}{x^2\text{Li}_{\lambda-1}(e^{-1/\kappa})}.
\end{equation}

\begin{equation}
	G_1''(x)=\frac{\text{Li}_{\lambda-3}(xe^{-1/\kappa})-3\text{Li}_{\lambda-2}(xe^{-1/\kappa})+2\text{Li}_{\lambda-1}(xe^{-1/\kappa})}{x^3\text{Li}_{\lambda-1}(e^{-1/\kappa})}.
\end{equation} 

Inserting Eq. (S37-41) into Eq. (S5-13), we can get $f_\text{b}$, $s_\text{GCC}$, $s_\text{BCC}$, $\langle B \rangle$ and $\text{var}(B)$. Results are shown in Fig. \ref{fig:1},\ref{fig:2}.

\begin{figure} 
\centering
\includegraphics[width=0.48\textwidth]{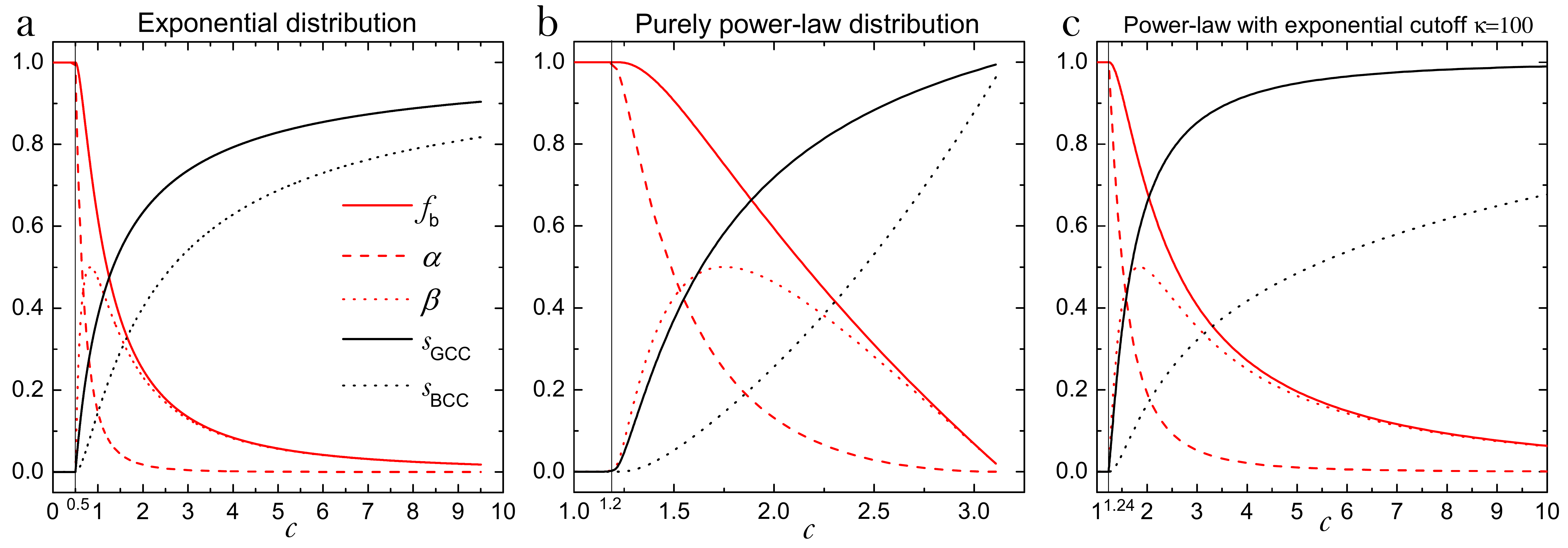}
\caption{\label{fig:1} \textbf{The analytical results of the bridge fraction $f_\text{b}$, relative sizes of the BCC and GCC in different random graphs.} (\textbf{a}) Exponentially distributed graphs. (\textbf{b}) Purely power-law distributed graphs. 
(\textbf{c}) Power-law distribution with exponential cutoff parameter $\kappa=100$.}
\end{figure}

\begin{figure}
\centering
\includegraphics[width=0.48\textwidth]{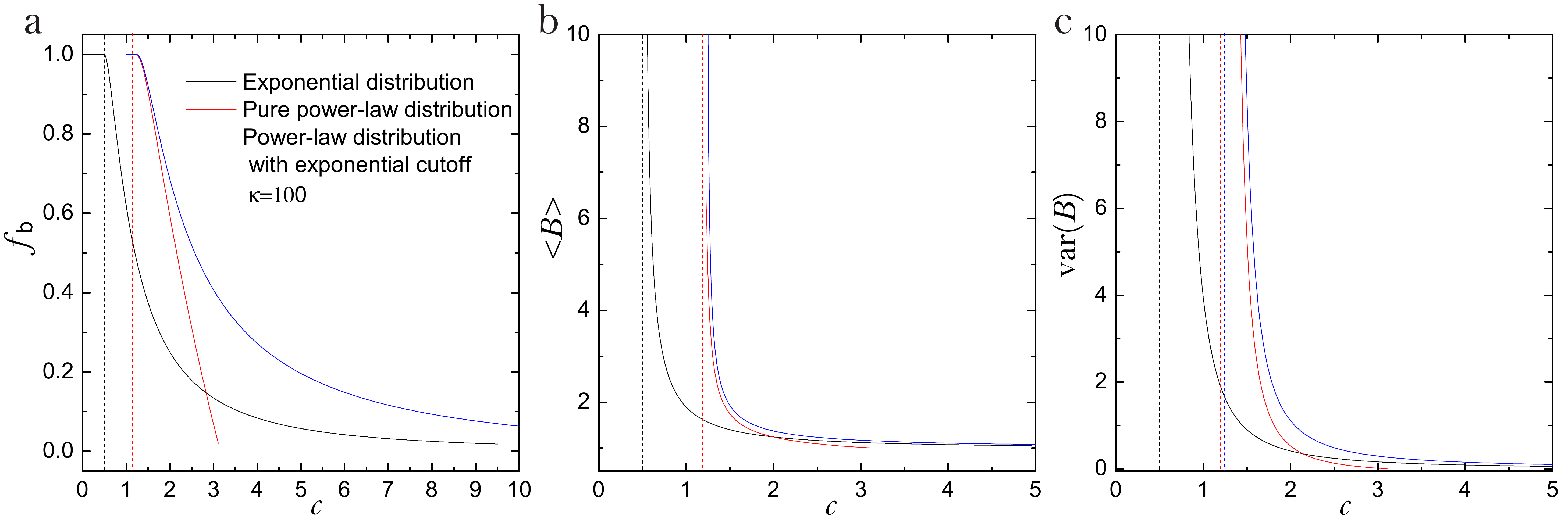}
\caption{\label{fig:2} \textbf{Bridge fraction, average and variance of bridgeness in different degree distribution.} All the results are analytical and the dashed lines mark their percolation positions.}
\end{figure}

\subsection{Static model}
In the main text, we use static model to generate scale-free (SF) random graphs \cite{goh2001universal}. This model consists of following steps \cite{liu2012core}:
\begin{itemize}
	\item Given $N$ isolated nodes, we label them from $1$ to $N$. For each node i, we assign a weight $p_i \propto i^{-a}$, where $a=\frac{1}{\lambda-1}$ and $\lambda$ is the characteristic parameter of SF graphs. 
    \item Then we randomly choose two nodes according to their weights and connect them if they are not connected. Self-links and multi-links are forbidden here. We repeat this step until $M=cN/2$ links are added.
\end{itemize}

The degree distribution of the static mode can be analytically derived as \cite{catanzaro2005analytic,lee2006intrinsic}:
  \begin{equation}
	 P(k)=\frac{[c(1-a)]^{1/a}}{a}\frac{\Gamma(k-1/a,c(1-a))}{\Gamma(k+1)},
\end{equation}
with $\Gamma(s)$ the gamma function and $\Gamma(s.x)$ the upper incomplete gamma function. When $k \to \infty$, $P(k) \sim k^{-(1+1/a)}=k^{-\lambda}$. Therefore, we can build different SF random graphs by tuning $a$.
The generating functions are:
  \begin{equation}
	 G_0(x)=\frac{1}{a}E_{1+\frac{1}{a}}[c(1-a)(1-x)],
 \end{equation}

  \begin{equation}
	 G_1(x)=\frac{1-a}{a}E_{\frac{1}{a}}[c(1-a)(1-x)],
 \end{equation}
where $E_n =\int_1^{\infty} e^{-xy}y^{-n}\mathrm{d}y$ is the exponential integral. Note that the derivative of $E_n$ follows $E_n'=-E_{n-1}(x)$.
From the generating functions, we can derive $f_\text{b}$, $s_\text{GCC}$, $s_\text{BCC}$, $\langle B \rangle$ and $\text{var}(B)$. Results are shown in main text Fig. 5.

\section{Network datasets}
Detailed information about the real-world networks analyzed in this paper are listed in Tables S1 with brief descriptions. We categorize networks according to their types and show their names, numbers of nodes, edges, bridges and biconnected components, size of the GCC as well as the average, variance and maximum of bridgeness.
\begin{sidewaystable}[b]
\caption{Real network datasets.\label{real}}
\centering
\begin{tabular}{l l l l l c l c l c l}
\hline
\hline
category & name & $N$ & $L$ & $s_\text{GCC}$ & $N_\text{BCC}$ & $N_\text{bridges}$ & $\langle B \rangle$ & $\text{var}(B)$ & $B_\text{max}$ & description \\
\hline
Air traffic & USairport500 \cite{colizza2007reaction}& 500 & 2,980 & 500 & 4 & 82 & 1.15 & 0.37 & 5 & Network among the top 500 busiest \\ 
 & & & & & & & & & & commercial airports in US\\
	& USairport-2010 \cite{opsahl2011anchorage}& 1,574 & 17,215 & 1,572 & 1 & 335 & 1.02 & 0.023 & 2 & US airport network in 2010\\
    	& openfights \cite{opsahl2011anchorage}& 2,939 & 15,677 & 2,905 & 6 & 724 & 1.15 & 0.89 & 20 &  non-US-based airport network\\
\hline
Road networks \cite{leskovec2015snap}& RoadNet-CA & 1,965,206 & 2,766,607 & 1,957,027 & 4,042 & 376,517 & 1.598 & 6.53 & 162 &  California road network\\
	& RoadNet-PA & 1,088,092 & 1,541,898 & 1,087,562 & 1,815 & 216,994 & 1.39 & 2.16 & 94 & Pennsylvania road network\\
    	& RoadNet-TX & 1,379,917 & 1,921,660 & 1,351,137 & 3,054 & 290,333 & 1.48 & 5.32 & 209 &   Texas road network\\
\hline
Power grids (PG) & PG-Texas \cite{bianconi2008local}& 4889 & 5855 & 4889 & 3 & 1425 & 1.40 & 1.61 & 27 & Power grid in Texas\\
	& PG-WestState \cite{watts1998collective}& 4,941 & 6,594 & 4,941 & 16 & 1,611 & 1.58 & 2.37 & 18 & High-voltage power grid in the Western\\ 
  & & & & & & & & & & states in US\\
\hline
Internet: & as20000102 & 6,474 & 12,572 & 6,474 & 1 & 2,451 & 1.04 & 0.064 & 5 & AS graph from January 02, 2000\\
Autonomous & oregon1-010331 & 10,670 & 22,002 & 10,670 & 1 & 3,799 & 1.03 & 0.077 & 10 &  AS peering information inferred from\\
    & & & & & & & & & &  Oregon route-views (I)\\
Systems(AS) \cite{leskovec2015snap} & oregon1-010407 & 10,729 & 21,999 & 10,729 & 1 & 3,848 & 1.03 & 0.048 & 5 & Same as above (at different time)\\
 & oregon1-010414 & 10,790 & 22,469 & 10,790 & 1 & 3,853 & 1.03 & 0.048 & 5 & Same as above (at different time)\\
  & oregon1-010421 & 10,859 & 22,747 & 10,859 & 1 & 3,855 & 1.03 & 0.047 & 6 &  Same as above (at different time)\\
   & oregon1-010428 & 10,886 & 22,493 & 10,886 & 1 & 3,844 & 1.03 & 0.049 & 7 & Same as above (at different time)\\
    & oregon1-010505 & 10,943 & 22,607 & 10,943 & 1 & 3,832 & 1.03 & 0.046 & 6 & Same as above (at different time)\\    
\hline
\end{tabular}
\end{sidewaystable}

\begin{sidewaystable}
\centering
\begin{tabular}{l l l l l c l c l c l}
\hline
\hline
category & name & $N$ & $L$ & $s_\text{GCC}$ & $N_\text{BCC}$ & $N_\text{bridges}$ & $\langle B \rangle$ & $\text{var}(B)$ & $B_\text{max}$ & description \\
\hline
Internet:     & oregon1-010512 & 11,011 & 22,677 & 11,011 & 1 & 3,909 & 1.03 & 0.047 & 7 & Same as above (at different time)\\
Autonomous & oregon1-010519 & 11,051 & 22,724 & 11,051 & 1 & 3,920 & 1.03& 0.044 & 6 & Same as above (at different time)\\
Systems(AS) \cite{leskovec2015snap} & oregon1-010526 & 11,174 & 23,409 & 11,174 & 1 & 3,946 & 1.02 & 0.035 & 6 & Same as above (at different time)\\
 & oregon2-010331 & 10,900 & 31,180 & 10,900 & 1 & 3,274 & 1.02 & 0.025 & 6 & 
 AS peering information inferred from Oregon\\ 
     & & & & & & & & & & route-views (II)\\ 
 & oregon2-010407 & 10,981 & 30,855 & 10,981 & 1 & 3,332 & 1.02 & 0.032 & 5 & Same as above (at different time)\\
 & oregon2-010414 & 11,019 & 31,761 & 11,019 & 1 & 3,316 & 1.014 & 0.019 & 4 & Same as above (at different time)\\
 & oregon2-010421 & 11,080 & 31,538 & 11,080 & 1 & 3,294 & 1.015 & 0.022 & 5 &Same as above (at different time)\\
 & oregon2-010428 & 11,113 & 31,434 & 11,113 & 1 & 3,283 & 1.016 & 0.020 & 3 & Same as above (at different time)\\
  & oregon2-010505 & 11,157 & 30,943 & 11,157 & 1 & 3,282 & 1.014 & 0.017 & 3 & Same as above (at different time)\\
   & oregon2-010512 & 11,260 & 31,303 & 11,260 & 2 & 3,296 & 1.017 & 0.026 & 5 & Same as above (at different time)\\
    & oregon2-010519 & 11,375 & 32,287 & 11,375 & 1 & 3,320 & 1.016 & 0.019 & 3 & Same as above (at different time)\\
     & oregon2-010526 & 11,461 & 32,730 & 11,461 & 1 & 3,351 & 1.015 & 0.018 & 4 & Same as above (at different time)\\
 \hline
Electronic & s838 & 512 & 819 & 512 & 1 & 49 & 1.33 & 0.22 & 2 & Electronic sequential logic circuit\\
circuits \cite{milo2002network} & s420 & 252 & 399 & 252 & 1 & 25 & 1.32 & 0.22 & 1 & Same as above\\   
  & s208 & 122 & 189 & 122 & 1 & 13 & 1.31 & 0.21 & 2 & Same as above\\ 
\hline
World Wide Web & stanford.edu \cite{leskovec2015snap}& 281,903 & 1,992,636 & 255,265 & 1,073 & 27,344 & 4.42 & 2,081.1 & 4907 & WWW from stanford.edu domain\\
(WWW) & nd.edu \cite{albert1999internet}& 325,729 & 1,090,108 & 325,729 & 308 & 166,376 & 1.58 & 264.29 & 2,660 & WWW from nd.edu domain\\ 
\hline
Neural network \cite{watts1998collective} & \emph{C.elegans} & 297 & 2,148 & 297 & 1 & 15 & 1.00 & 0.00 & 0 & Neural network of \emph{C.elegans}\\
\hline
\end{tabular}
\end{sidewaystable}

\begin{sidewaystable}
\centering
\begin{tabular}{l l l l l c l c l c l}
\hline
\hline
category & name & $N$ & $L$ & $s_\text{GCC}$ & $N_\text{BCC}$ & $N_\text{bridges}$ & $\langle B \rangle$ & $\text{var}(B)$ & $B_\text{max}$ & description \\
\hline
Inertnet: & p2p-Gnutella04 & 10,876 & 39,994 & 10,876 & 1 & 2,497 & 1.020 & 0.099 & 12 & Gnutella p2p file sharing network\\
peer-to-peer (p2p)	& p2p-Gnutella05 & 8,846 & 31,839 & 8,842 & 1 & 2,009 & 1.011 & 0.012 & 3 &  Same as above (at different time)\\
(p2p)file sharing & p2p-Gnutella06 & 8,717 & 31,525 & 8,717 & 1 & 1,990 & 1.012 & 0.013 & 3 & Same as above (at different time)\\
file sharing networks \cite{leskovec2015snap} & p2p-Gnutella08 & 6,301 & 20,777 & 6,299 & 1 & 1,765 & 1.016 & 0.067 & 10 &  Same as above (at different time)\\
 &p2p-Gnutella09 & 8,114 & 26,013 & 8,104 & 1 & 2,503 & 1.008 & 0.010 & 2 &  Same as above (at different time)\\
  	& p2p-Gnutella24 & 26,518 & 65,369 & 26,498 & 1 & 10,989 & 1.005 & 0.0053 & 2 & Same as above (at different time)\\
    	& p2p-Gnutella25 & 22,687 & 54,705 & 22,663 & 1 & 9,322 & 1.006 & 0.014 & 9 &  Same as above (at different time)\\
        	& p2p-Gnutella30 & 36,682 & 88,328 & 36,646 & 1 & 16,477 & 1.004 & 0.005 & 3 &  Same as above (at different time)\\
            	& p2p-Gnutella31 & 62,586 & 147,892 & 62,561 & 1 & 28,759 & 1.005 & 0.0075 & 6 &  Same as above (at different time)\\
\hline
Food webs (FW) & baydry \cite{ulanowicz1998fy97} & 128 & 2,106 & 128 & 1 & 0 & 0.00 & 0.00 & 0 & FW at Florida Bay, Dry Season\\
 & baywet \cite{ulanowicz1998fy97}& 128 & 2,075 & 128 & 1 & 0 & 0.00 & 0.00 & 0 & FW at Florida Bay, Wet Season\\
  & Chesapeake \cite{baird1989seasonal}& 39 & 170 & 39 & 1 & 0 & 0.00 & 0.00 & 0 & FW at Chesapeake Bay Mesohaline Net\\
   & ChesLower \cite{hagy2002eutrophication}& 37 & 167 & 37 & 1 & 0 & 0.00 & 0.00 & 0 & FW at Lower Chesapeake Bay in Summer\\
    & ChesMiddle \cite{hagy2002eutrophication}& 37 & 198 & 37 & 1 & 0 & 0.00 & 0.00 & 0 & FW at Middle Chesapeake Bay in Summer\\
     & ChesUpper \cite{hagy2002eutrophication}& 37 & 199 & 37 & 1 & 0 & 0.00 & 0.00 & 0 & FW at Upper Chesapeake Bay in Summer\\
      & CrystalC \cite{ulanowicz2012growth}& 24 & 114 & 24 & 1 & 0 & 0.00 & 0.00 & 0 & FW at Crystal River Creek (Control)\\
 & CrystalD \cite{ulanowicz2012growth}& 24 & 92 & 24 & 1 & 0 & 0.00 & 0.00 & 0 & FW at Crystal River Creek (Delta Temp)\\
  & cypdry \cite{websitepajek}& 71 & 618 & 71 & 1 & 0 & 0.00 & 0.00 & 0 & FW at Cypress, Dry Season\\
  & cypwet \cite{websitepajek}& 71 & 612 & 71 & 1 & 0 & 0.00 & 0.00 & 0 & FW at Cypress, Wet Season\\
\hline
\end{tabular}
\end{sidewaystable}

\begin{sidewaystable}
\centering
\begin{tabular}{l l l l l c l c l c l}
\hline
\hline
category & name & $N$ & $L$ & $s_\text{GCC}$ & $N_\text{BCC}$ & $N_\text{bridges}$ & $\langle B \rangle$ & $\text{var}(B)$ & $B_\text{max}$ & description \\
\hline
Food webs (FW) & Everglades \cite{ulanowicz2000annual}& 69 & 880 & 69 & 1 & 0 & 0.00 & 0.00 & 0 & FW at Everglades Graminoid Marshes\\
 & Florida \cite{ulanowicz1998fy97}& 128 & 2,075 & 128 & 1 & 0 & 0.00 & 0.00 & 0 & FW at Florida Bay Trophic Exchange Matrix\\
  & gramdry \cite{ulanowicz2000annual}& 69 & 879 & 69 & 1 & 0 & 0.00 & 0.00 & 0 & FW at Everglades Graminoids, Dry Season\\
   & gramwet \cite{ulanowicz2000annual}& 69 & 880 & 69 & 1 & 0 & 0.00 & 0.00 & 0 & FW at Everglades Graminoids, Wet Season\\
    & grassland \cite{dunne2002two}& 88 & 137 & 88 & 4 & 30 & 1.80 & 2.69 & 5 & FW at Grassland\\
     & littlerock \cite{martinez1991artifacts}& 183 & 2,434 & 183 & 1 & 1 & 1.00 & 0.00 & 0 & FW at Little Rock lake.\\
      & mangdry \cite{patricio2005thesis}& 97 & 1,445 & 97 & 1 & 0 & 0.00 & 0.00 & 0 & FW at Mangrove Estuary, Dry Season\\
 & mangwet \cite{patricio2005thesis}& 97 & 1,446 & 97 & 1 & 0 & 0.00 & 0.00 & 0 & 
FW at Mangrove Estuary, Wet Season\\
  & Maspalomas \cite{almunia1999benthic}& 24 & 77 & 24 & 1 & 0 & 0.00 & 0.00 & 0 & 
FW at Charca de Maspalomas\\
   & Michigan \cite{websitepajek}& 39 & 209 & 39 & 1 & 0 & 0.00 & 0.00 & 0 & FW at Lake Michigan Control network\\
  & Mondego \cite{patricio2005thesis}& 46 & 358 & 46 & 1 & 1 & 1.00 & 0.00 & 0 & FW at Mondego Estuary - Zostrea site\\
   & Narragan \cite{patricio2005thesis}& 35 & 204 & 35 & 1 & 0 & 0.00 & 0.00 & 0 & FW at Narragansett Bay Model\\
    & Rhode \cite{websitepajek}& 20 & 45 & 19 & 1 & 0 & 0.00 & 0.00 & 0 & FW at Rhode River Watershed - Water Budget\\
     & seagrass \cite{christian1999organizing}& 49 & 223 & 49 & 1 & 0 & 0.00 & 0.00 & 0 & FW at St. Marks Seagrass.\\
      & silwood \cite{memmott2000predators}& 154 & 365 & 153 & 1 & 60 & 1.20 & 0.56 & 4 &  FW at Silwood Park\\
 & StMarks \cite{baird1998assessment}& 54 & 350 & 54 & 1 & 0 & 0.00 & 0.00 & 0 & 
FW at St. Marks River (Florida) Flow network\\
  & stmartin \cite{goldwasser1993construction}& 45 & 224 & 45 & 1 & 2 & 1.00 & 0.00 & 1 & FW at St. Martin Island\\
       & ythan \cite{dunne2002two}& 135 & 596 & 135 & 1 & 7 & 1.00 & 0.00 & 1 & FW at Ythan Estuary\\
\hline
\end{tabular}
\end{sidewaystable}

\begin{sidewaystable}
\centering
\begin{tabular}{l l l l l c l c l c l}
\hline
\hline
category & name & $N$ & $L$ & $s_\text{GCC}$ & $N_\text{BCC}$ & $N_\text{bridges}$ & $\langle B \rangle$ & $\text{var}(B)$ & $B_\text{max}$ & description \\
\hline
Transcriptional & TRN-Yeast-Babu \cite{balaji2006comprehensive}& 4,441 & 12,864 & 4,441 & 1 & 1,556 & 1.00 & 0.00 & 0 & Transcriptional regulatory network \\
    & & & & & & & & & & of \emph{S.cerevisiae}\\
regulatory networks & TRN-Yeast-Alon \cite{milo2002network}& 688 & 1,078 & 662 & 1 & 343 & 1.31 & 3.17 & 20 & Same as above (compiled by different\\
    & & & & & & & & & &  group)\\
(TRN)  & TRN-EC-RDB64 \cite{gama2008regulondb}& 1,550 & 3,234 & 1,454 & 1 & 616 & 1.046 & 0.18 & 7 & Transcriptional regulatory network\\
    & & & & & & & & & &  of \emph{E.coli}\\
   & TRN-EC-Alon \cite{milo2002network}& 418 & 519 & 328 & 3 & 251 & 1.45 & 5.86 & 24 & Same as above (compiled by different\\
       & & & & & & & & & &  group)\\
\hline
Communication & Cellphone \cite{song2010limits}& 36,595 & 56,853 & 30,420 & 415 & 13,382 & 1.66 & 2.56 & 36 & Call network of cell phone users\\
networks & Email-Enron \cite{leskovec2015snap}& 36,692 & 183,831 & 33,696 & 366 & 10,714 & 1.08 & 1.53 & 105 &  Email communication network from\\
       & & & & & & & & & &  Enron\\
 & Email-EuAll \cite{leskovec2015snap}& 265,214 & 364,481 & 224,832 & 10 & 213,190 & 1.012 & 0.52 & 169 &  Email network from a large European \\ 
    & & & & & & & & & & research institute\\
 & Email-epoch \cite{eckmann2004entropy}& 3,188 & 31,857 & 3,186 & 1 & 613 & 1.0016 & 0.0016 & 1 &  Email network in a university\\
  & UCIonline \cite{opsahl2009clustering}& 1,899 & 13,838 & 1,893 & 1 & 398 & 1.020 & 0.025 & 2 &  Online message network of students\\
     & & & & & & & & & &  at UC, Irvine\\
   & WikiTalk \cite{leskovec2015snap}& 2,394,385 & 4,659,565 & 2,388,953 & 13 & 1,768,844 & 1.0024 & 0.0043 & 19 &  Wikipedia talk network\\
 \hline
Social networks & Epinions \cite{richardson2003semantic}& 75,888 & 405,740 & 75,877 & 87 & 38,679 & 1.12 & 0.35 & 34 & Who-trusts-whom network of\\
    & & & & & & & & & & Epinions.com\\
\hline
\end{tabular}
\end{sidewaystable}

\begin{sidewaystable}
\centering
\begin{tabular}{l l l l l c l c l c l}
\hline
\hline
category & name & $N$ & $L$ & $s_\text{GCC}$ & $N_\text{BCC}$ & $N_\text{bridges}$ & $\langle B \rangle$ & $\text{var}(B)$ & $B_\text{max}$ & description \\
\hline
Social networks & college student \cite{van2003evolution,milo2004superfamilies}& 32 & 80 & 32 & 1 & 0 & 0.00 & 0.00 & 0 & Social networks of positive sentiment \\
 & prison inmate \cite{van2003evolution,milo2004superfamilies}& 67 & 142 & 67 & 1 & 5 & 1.20 & 0.16 & 1 &  Same as above (prison inmates) \\
 & Slashdot-1 \cite{leskovec2015snap}& 77,357 & 468,554 & 77,350 & 46 & 29,680 & 1.063 & 0.096 & 8 & Slashdot social network\\
 & Slashdot-2 \cite{leskovec2015snap}& 81,871 & 497,672 & 81,867 & 46 & 29,998 & 1.066 & 0.1084 & 9 & Same as above (at different time)\\
 & Slashdot-3 \cite{leskovec2015snap}& 82,144 & 500,481 & 82,140 & 46 & 30,086 & 1.066 & 0.1084 & 8 & Same as above (at different time)\\
 & Slashdot-4 \cite{leskovec2015snap}& 77,360 & 469,180 & 77,360 & 46 & 29,649 & 1.063 & 0.096 & 9 & Same as above (at different time)\\
 & Slashdot-5 \cite{leskovec2015snap}& 82,168 & 504,230 & 82,168 & 45 & 30,035 & 1.066 & 0.108 & 8 & Same as above (at different time)\\
 & Twitter \cite{leskovec2015snap}& 81,306 & 1,342,296 & 81,306 & 1 & 4,996 & 1.00 & 0.00 & 0 &  Social circles from Twitter\\
 & WikiVote \cite{leskovec2015snap}& 7,115 & 100,762 & 7,066 & 1 & 2,306 & 1.006 & 0.011 & 5 &   Wikipedia who-votes-on-whom network\\
 & Youtube \cite{leskovec2015snap}& 1,134,890 & 2,987,624 & 1,134,890 & 1,944 & 667,090 & 1.20 & 1.25 & 91 & Youtube online social network\\
 \hline
Intra- & Freemans-1 \cite{freeman1979networkers}& 34 & 415 & 34 & 1 & 0 & 0.00 & 0.00 & 0 &  Social network of network researchers\\
organizational & Freemans-2 \cite{freeman1979networkers}& 34 & 474 & 34 & 1 & 0 & 0.00 & 0.00 & 0 &Same as above (at different time)\\
networks & Freemans-3 \cite{freeman1979networkers}& 32 & 266 & 32 & 1 & 0 & 0.00 & 0.00 & 0 & Same as above (at different time)\\
 & Consulting-1 \cite{cross2004hidden}& 46 & 550 & 46 & 1 & 0 & 0.00 & 0.00 & 0 & Social network from a consulting company\\
  & Consulting-2 \cite{cross2004hidden}& 46 & 505 & 46 & 1 & 0 & 0.00 & 0.00 & 0 & Same as above (different evaluation)\\
   & Manufacturing-1 \cite{cross2004hidden}& 77 & 1,344 & 77 & 1 & 0 & 0.00 & 0.00 & 0 &  Social network from a manufacturing\\
 & & & & & & & & & &  company\\
    & Manufacturing-2 \cite{cross2004hidden}& 77 & 1,341 & 77 & 1 & 0 & 0.00 & 0.00 & 0 & Same as above (different evaluation)\\
 \hline
PPI networks& CCSB-Y2H \cite{yu2008high}& 892 & 1,015 & 787 & 1 & 627 & 1.145 & 0.197 & 4 & Interactions from CCSB-YI11\\
\hline
\end{tabular}
\end{sidewaystable}

\begin{sidewaystable}
\centering
\begin{tabular}{l l l l l c l c l c l}
\hline
\hline
category & name & $N$ & $L$ & $s_\text{GCC}$ & $N_\text{BCC}$ & $N_\text{bridges}$ & $\langle B \rangle$ & $\text{var}(B)$ & $B_\text{max}$ & description \\
\hline
PPI networks   & Y2H-union \cite{yu2008high}& 1,304 & 1,412 & 946 & 1 & 820 & 1.48 & 1.36 & 12 & The union of CCSB-YI1, Ito-core and\\
    & & & & & & & & & &  Uetz-screen\\
 & wi2004 \cite{simonis2009empirically} & 743 & 827 & 511 & 1 & 477 & 1.37 & 0.91 & 9 & Interactions from 2004 Y2H screen\\
 & wi2007 \cite{simonis2009empirically} & 814 & 863 & 545 & 1 & 522 & 1.40 & 0.97 & 8 & Interactions from 2007 Y2H screen\\
 & HI-I-05 \cite{rual2005towards} & 1,551 & 2,612 & 1,307 & 2 & 803 & 1.20 & 0.41 & 8 & Proteome-scale map of the human binary interactome\\
    & & & & & & & & & &  network generated by systematically screening Space-I\\
 & HI-II-14 \cite{rolland2014proteome}& 4,305 & 13,428 & 4,100 & 1 & 1,652 & 1.08 & 0.13 & 4 & Same as above (by different screening space)\\
\hline
\end{tabular}
\end{sidewaystable}

%
\putbib
\end{bibunit}

\end{document}